# Software Engineering in Australasia

Sherlock A. Licorish[1], Christoph Treude[2], John Grundy[3], Chakkrit Tantithamthavorn[3], Kelly Blincoe[4], Stephen MacDonell[1,5], Li Li[3], Jean-Guy Schneider[6]
*[1]University of Otago, New Zealand, {sherlock.licorish, stephen.macdonell}@otago.ac.nz*
*[2]University of Adelaide, Australia, christoph.treude@adelaide.edu.au*
*[3]Monash University, Australia, {john.grundy, chakkrit, li.li}@monash.edu*
*[4]University of Auckland, New Zealand, k.blincoe@auckland.ac.nz*
*[5]Auckland University of Technology, New Zealand, stephen.macdonell@aut.ac.nz*
*[6]Deakin University, Australia, jeanguy.schneider@deakin.edu.au*

## 1. INTRODUCTION

Six months ago an important call was made for researchers globally to provide insights into the way Software Engineering is done in their region. Heeding this call, we hereby outline the position Software Engineering in Australasia (New Zealand and Australia). This article first considers the software development methods, practices and tools that are popular in the Australasian software engineering community. We then briefly review the particular strengths of software engineering researchers in Australasia. Finally, we make an open call for collaborators by reflecting on our current position and identifying future opportunities.

## 2. SOFTWARE DEVELOPMENT METHODS, PRACTICES AND TOOLS

Historically, much of what we know about software development methods, practices, and tools is shaped by technology vendors. In many cases, these vendors are positioned in very large economies, and, thus, those software engineering communities can be equally large. For instance, 53% of the software development companies represented in the popular VersionOne 2014 survey had > 1000 employees[1], in contrast to the small and medium-sized enterprises (SMEs) that are typical in Australasia. Unsurprisingly, 65% of those surveyed by VersionOne were from North America, and while there has been improvement in participation from other continents in 2020, the recent results from VersionOne are still heavily skewed to North America and Europe[2]. From time to time however, the utility of these methods, practices and tools is investigated beyond these larger economies, providing lessons for the software engineering communities in other regions.

One such study surveyed practitioners working in Brazil, Finland, and New Zealand in a transnational study where similar findings were reported across the three territories [1]. Outcomes from this study revealed that the sample of practitioners surveyed focused on a small portfolio of projects that were of short duration. In addition, Scrum and Kanban were used most; however, some practitioners also used more traditional methods. Coding Standards, Simple Design and Refactoring were used most by practitioners, and these practices were held to be largely suitable for project and process management [1]. Interestingly, another study also found the same threats to software development success across Brazil, Finland, and New Zealand [2]. While evidence here suggests that there may be similarities in the software development landscapes across the three countries, and particularly in terms of the size of organisations, this evidence may also point to the global language that is 'software engineering'. In fact, globally, software engineers now place increasing value on agility and methods and practices that aid this process, notwithstanding the various hybrids that are typically implemented in practice [3].

This is fitting, given that sales of software and related services rose by 31% from 2017 to 2019 in New Zealand to $9.8 billion, with published software accounting for $3.1 billion[3]. The positive trend is also reported in Australia; between 2005 and 2019, the productivity benefits from the growing digital economy increased Australia's steady state GDP per capita by 6.5%[4]. With the adoption of 5G and other emerging digital technologies, the contribution of things digital to productivity will only grow. The Australian software industry landscape is responding to this growth with diversity, ranging from vibrant startups to Sydney-based tech giant Atlassian, known for products such as Jira, Confluence, Bitbucket, and Trello. According to recent work on analysing requirements changes in Agile teams [4], the most frequently practised Agile method in

[1] https://explore.digital.ai/state-of-agile/9th-annual-state-of-agile-report
[2] https://explore.digital.ai/state-of-agile/14th-annual-state-of-agile- report
[3] https://www.stats.govt.nz/information-releases/information-and-communication-technology-supply-survey-2019
[4] https://www2.deloitte.com/au/en/pages/economics/articles/australias-digital-pulse.html

Australian companies is Scrum, with a mean team size of 18 members and a mean iteration length of just under five weeks. The most followed Agile techniques are daily standups, Kanban boards, sprints, and the sprint backlog, whereas pair programming and self- assignment were the least followed practices.

## 3. SOFTWARE DEVELOPMENT RESEARCH

Australasian software engineering researchers work in all knowledge areas. Figure 1 provides a word cloud depicting Australasian software engineering researchers' focus, where the size of the word denotes the relative degree of research focus across all active researchers. Of the areas listed, empirical software engineering, formal methods, cloud computing, and artificial intelligence are particular strengths. Other significant research areas cover specific variances of cloud computing, formal methods, artificial intelligence and machine learning, programming languages and algorithms, program analysis and repair, mining software repositories, software security, software engineering tool support and HCI, requirements engineering, software design and architecture, software engineering education, software quality and testing, software development methods and practices, process mining, software measurements, empirical software engineering, human factors in software engineering, green software engineering, NLP applications, distributed systems, empirical methods and data quality, theory building in software engineering, project management, and so on.

Our research outcomes have also led to impact outside of academia, underscoring the prominence of some of the topics in Figure 1. For instance, Chakkrit Tantithamthavorn's work on developing analytics for predicting wait-time in the emergency department is currently deployed in many hospitals in Australia[5].

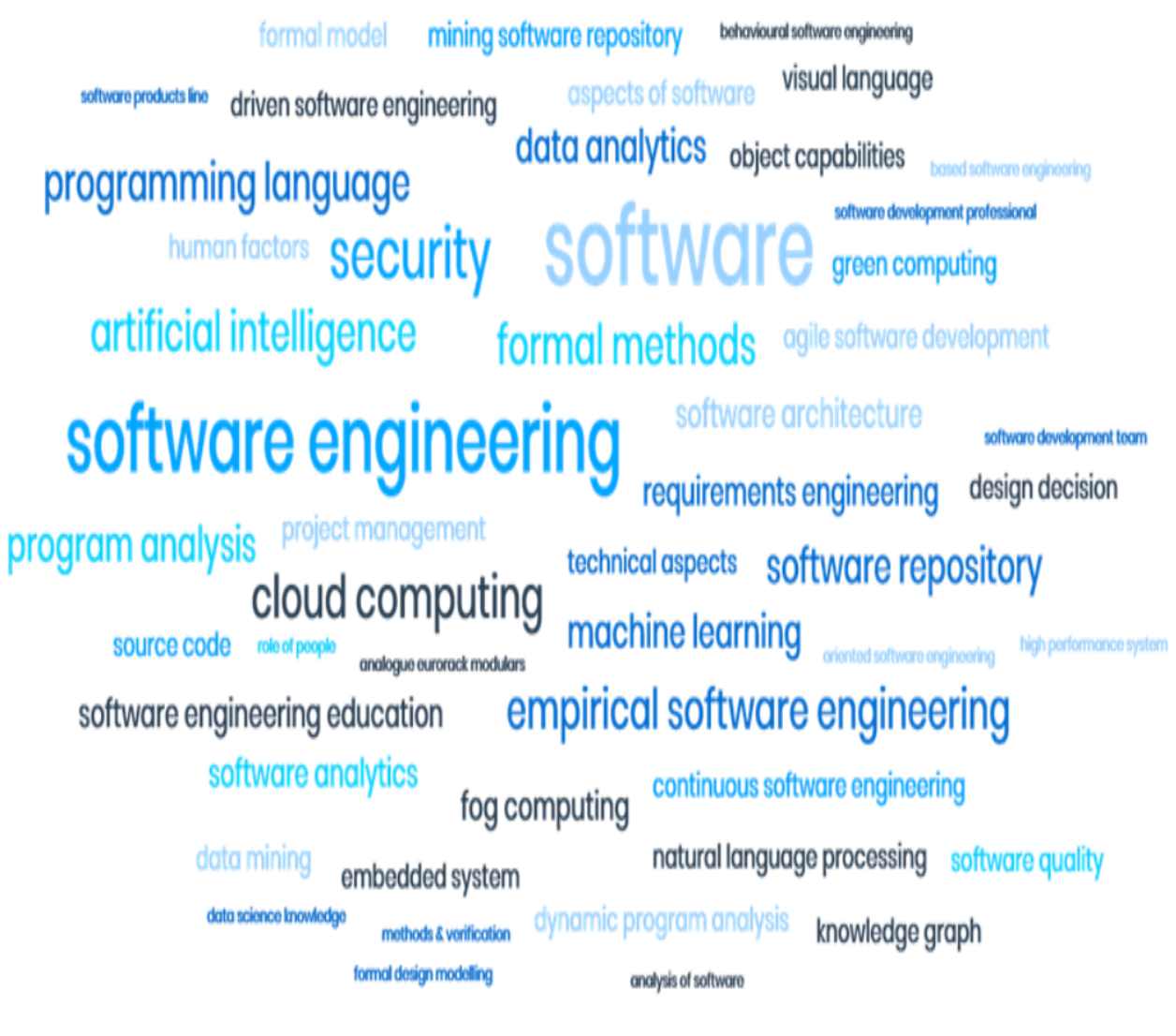

Figure 1. Australasian software engineering researchers' focus

[5] https://www.monash.edu/it/about-us/news-and-events/latest/articles/2020/how-data-helps-doctors-make-life-saving-decisions-in-emergency-care

## 4. REFLECTIONS AND FUTURE OPPORTUNITIES

Software engineering researchers across Australasia continue to participate actively in every area of research worldwide, and we are open for business. Across the spectrum of the esteem metrics, Australasian software engineering researchers continue to perform admirably, including with scientific acknowledgments and awards, editorial tasks, organising roles in academic events, PC memberships, peer review records, external refereeing (including for major grants), invited talks and lectures, research visits, and professional memberships. For instance, APSEC2016, EASE2018, ICSME2020, VL/HCC2020, and ASE2020 were all recently held in Australasia. ASE2021, RE2022, and ICSE2023 will be held in Melbourne, Australia. Software engineering researchers have also increasingly received funding; for instance, software engineering academics in New Zealand have received several research grant awards from the New Zealand government in recent years[6], including a prestigious Marsden award which was received by Kelly Blincoe. Notable Science for Technological Innovation National Science Challenge (SfTI) funding was also awarded to Michael Homer, Steve Reeves, Sherlock Licorish and Amjed Tahir. Furthermore, Panos Patros is part of a team awarded $12.5M to decarbonize the New Zealand Industrial Sector via Self-Adaptive Software.

Much of the recent software engineering research funding success in New Zealand may be attributed to the formation of Software Innovation New Zealand (SI^NZ) in October 2016. The goal of this committee is to bring together leading software engineering research teams from New Zealand universities in a coordinated centre of research excellence with a strong industry focus and raise the level and profile of New Zealand software engineering research. SI^NZ has also facilitated increased collaboration among software engineering researchers in New Zealand.

In Australia, Professor John Grundy has been awarded an ARC Laureate Fellowship, one of the most prestigious research awards in Australia. Seven early-career academics received an Australian Research Council (ARC) Discovery Early Career Researchers Award (DECRA) over the last few years (i.e., Aldeida Aleti, Christoph Treude, Marcel Boehme, Li Li, Xin Xia, Chakkrit Tantithamthavorn and Patanamon Thongtanunam).

Notwithstanding the COVID-19 pandemic and the barriers to travel recently, we invite the world to travel to Australasia for research study leave to facilitate more exchanges. You may contact any of the authors of this article or others in the wider Australasian software engineering community about visiting. In addition, there is an open call for support for Australasia's bids for hosting conferences, and also for colleagues to travel to the region for conference attendance and to forge partnerships. In the short term, we warmly look forward to hosting you for ASE2021[7].

[6] https://softwareinnovation.nz/news/
[7] https://conf.researchr.org/home/ase-2021

## ACKNOWLEDGMENTS

We would like to thank ACM SIGSOFT and the instigators of the Software Engineering Worldwide column, especially Professors Marco Kuhrmann and Dietmar Pfahl, for the invitation to contribute to this article series.